\documentclass[pra,a4paper,floatfix,superscriptaddress]{revtex4}
\usepackage{pstricks,graphicx,amsmath,bbm,mathrsfs,amssymb,times,psfrag,pifont}
\usepackage{epsfig}
\def\PHAV{\hbox{\tiny PHAV}}
\def\DPHAV{\hbox{\tiny DPHAV}}
\def\BS{\hbox{\tiny BS}}
\def\d{\hbox{d}}
\def\a{\tilde{\alpha}}
\def\b{\tilde{\beta}}
\def\p{{\cal P}}
\def\GGamma{{C}}

\def\Id{{\boldsymbol 1}} 

\begin{document}

\title{Gaussian and Non-Gaussian operations on
non-Gaussian state: engineering non-Gaussianity}


\author{S.~Olivares}
\affiliation{Dipartimento di Fisica, Universit\`a degli Studi di Milano and CNISM, U.d.R. Milano Statale, via Celoria 16, I-20133 Milano, Italy.}

\author{A.~Allevi}
\affiliation{Dipartimento di Scienza e Alta Tecnologia, Universit\`a degli Studi dell'Insubria and CNISM, U.d.R. Como, via Valleggio 11, I-22100 Como, Italy}

\author{M.~Bondani}
\email{maria.bondani@uninsubria.it}
\affiliation{Istituto di Fotonica e Nanotecnologie, Consiglio Nazionale delle Ricerche, and CNISM, U.d.R. Como, via Valleggio 11, I-22100 Como, Italy}

\begin{abstract}
Multiple photon subtraction applied to a displaced phase-averaged coherent state, which is a non-Gaussian classical state, produces conditional states with a non trivial (positive) Glauber-Sudarshan $P$-representation. We theoretically and experimentally demonstrate that, despite its simplicity, this class of conditional states cannot be fully characterized by direct detection of photon numbers.
In particular, the non-Gaussianity of the state is a characteristics that must be assessed by phase-sensitive measurements.
We also show that the non-Gaussianity of conditional states can be manipulated by choosing suitable conditioning values and composition of phase-averaged states.
\end{abstract}
\maketitle

\section{Introduction}
In the last years we have witnessed an impressive step forward in the implementation of quantum information technologies, ranging from quantum communication to quantum computation. One of the main requirements to achieve this goal and pass from theoretical predictions to experimental realizations is the characterization of the quantum states and the operations involved in the protocols. Many efforts have been devoted to the introduction of new parameters aimed at characterizing the states: one of these is the non-Gaussianity. Here we show how such a parameter can be successfully used to characterize states which turn out to be experimentally indistinguishable from each other when setups based only on photon-number resolving (PNR) detectors and direct detection schemes are employed.
\par
Among the conditional measurements, photon subtraction (PS), both single and multiple, is an effective method to enhance quantum features of optical field states \cite{rev:PS}. The process, which is in general implemented by mixing at a beam splitter (BS) an input state with the vacuum, is obtained by performing PNR measurements at one output and selecting the other output only if a certain condition on the number of detected photons is satisfied. When PS is applied to nonclassical fields, it can lead to the generation of highly nonclassical states \cite{lamperti,perina13}, such as squeezed Fock states \cite{PS:gran,PS:oli} and cat-like states \cite{PS:cat,PS:cat:exp}. Furthermore, PS can be used in the continuos variable regime to enhance teleportation fidelity \cite{opatrny00,PS:coc,PS:tlp:oli} and non-locality \cite{PS:nha,PS:gar,PS:nl:inv,PS:allevi}.
\par
As a matter of fact, PS is in general a non-Gaussian operation: when applied to Gaussian states \cite{oli:rev}, namely states described by a Gaussian characteristic function, it generates conditional non-Gaussian states, whose characteristic functions are no longer Gaussian \cite{barbieri10,takeoka07}.
Based on this result, one would expect that applying PS to native non-Gaussian states would increase the amount of non-Gaussianity.
On the other hand, if one applies PS to native non-Gaussian states, one would expect an increase in the non-Gaussianity amount. This intuition is not always true and the effect of the PS operation is in general non-obvious.
\par
In this paper we investigate, both theoretically and experimentally, the effect of PS on a particular class of classical states, the displaced phase-averaged coherent states, which are obtained by first averaging the phase of a coherent state and then displacing it \cite{EPJD,curty:09}. These states are indeed useful candidates for our study, since they can be accurately generated, manipulated and characterized \cite{OE12}.
\par
The paper is structured as follows: in Sec.~\ref{s:phav} we summarize the statistical properties of phase-averaged coherent states, displaced or not, whereas in Sec.~\ref{s:cond:PNR} we describe the conditioning operations on these states by emphasizing the main features of the conditional states. Section~\ref{s:exp} presents the experimental setup used to generate such states.
The experimental reconstruction of the detected-photon distributions of conditional states is addressed in Sec.~\ref{s:cond:s}, where we also discuss the symmetry properties. Section~\ref{s:nonG} is devoted to the investigation of  non-Gaussianity in dependence on the different parameters characterizing our conditional states. Further discussions and concluding remarks are drawn in Sec.~\ref{s:concl}.


\section{Phase-averaged coherent states}\label{s:phav}
A phase-averaged coherent (PHAV) state, the main ingredient of our investigation, is obtained from a coherent state
\begin{equation}
| \beta \rangle = \exp\left(-\frac{|\beta|^2}{2}\right)
\sum_{k=0}^{\infty}\frac{\beta^k}{\sqrt{k!}}
| k \rangle,
\end{equation}
with $\beta = |\beta| e^{i\phi}\in{\mathbbm C}$ by averaging over the phase $\phi$.
A PHAV state with amplitude $\beta$ is described by a positive Glauber-Sudarshan $P$-representation \cite{glauber63,PHAV:rev}
\begin{equation}
\varrho_{\PHAV}(\beta) = \int_{\mathbbm C} \d^2 z\, P(z;\beta)\, |z \rangle \langle z |,
\end{equation}
where
\begin{equation}
P(z;\beta) = \frac{1}{2 \pi |\beta| }\,
\delta\left(|z|-|\beta|\right)
\end{equation}
and $\{|z \rangle \}$ is the basis of coherent states.
We can also expand the PHAV state on the photon-number basis, namely
\begin{equation}\label{dm}
\varrho_{\PHAV}(\beta) = \int_{-\pi}^{\pi} \frac{\d\phi}{2\pi}\, | \beta \rangle\langle
\beta | = \sum_{k=0}^{\infty} \p_k \left(\langle n \rangle\right)| k \rangle\langle k |,
\end{equation}
where
\begin{equation} \label{poisson}
\p_k(\langle n \rangle) = e^{-\langle n \rangle} \langle n \rangle^k / k!
\end{equation}
is the Poisson distribution, with mean value $\langle n \rangle=|\beta|^2$.
The latter representation is particularly useful to understand why a PHAV state alone is not suitable to produce conditional states by means of PS. In order to be an actual conditioning operation, PS requires the existence of intensity correlations between the two involved beams. In the case of classical states mixed with the vacuum at a BS, the amount of intensity correlations at the output is  a function of the first two moments of the photon-number statistics \cite{OLcorr}.
In particular, for a balanced BS we can write
\begin{equation} \label{gamma}
\GGamma = \frac{\sigma^2_n - \langle n \rangle}{\sigma^2_n + \langle n \rangle}
\end{equation}
$\langle n \rangle$ being the average number of photons of the state and
$\sigma^2_n$ the corresponding variance. In the presence of a Poisson photon-number distribution, as in the case of a coherent or a PHAV state, we have $\sigma^2_n = \langle n \rangle$ and no intensity correlations are observed between the two outputs, i.e. $\GGamma =0$. For this reason, the transmitted beam is unaffected by the operation performed on the reflected one and ${\it viceversa}$.
\par
On the other hand, if a displacement operation is applied to a PHAV state, the resulting state, namely the displaced PHAV (DPHAV) state, gives rise to two classically correlated beams when it is divided at a BS \cite{IJQI11}. In fact, its non-trivial photon-number distribution is super-Poissonian \cite{PHAV:rev}.
\par
If we start from the PHAV state given in Eq.~(\ref{dm}), the DPHAV state can be written as
\begin{align}\label{dm:D:phav}
\varrho_{\DPHAV}(\alpha,\beta) &= D(\alpha)\varrho_{\PHAV}(\beta) D^{\dag}(\alpha),
\nonumber\\[1ex]
&= \int_{\mathbbm C} \d^2 z\, P(z - \alpha;\beta)\, |z \rangle \langle z |,\nonumber\\[1ex]
  &=\int_{-\pi}^{\pi} \frac{\d\phi}{2\pi}\,
| \alpha + |\beta|\,e^{i \phi}\rangle\langle
\alpha + |\beta|\,e^{i \phi} |,
\end{align}
where $D(\alpha)=\exp(\alpha a^{\dag} - \alpha^* a)$ is the displacement operator, $a$ and $a^{\dag}$ are the annihilation and creation operators, respectively, $[a,a^{\dag}] = {\mathbbm I}$, and
\begin{equation}
P(z-\alpha;\beta) = \frac{1}{2 \pi |\beta| }\,
\delta\left(|z-\alpha|-|\beta| \right).
\end{equation}
It is worth noting that the Wigner function of a DPHAV state is still non-Gaussian, like in the case of PHAV states \cite{OLwigner,OE12}, but the state is phase-sensitive as it exhibits a non-diagonal density matrix in the photon-number basis.
\par
The photon-number distribution of DPHAV states can be written as (without loss of generality we can take $\alpha \in {\mathbbm R}$, $\alpha\ge 0$):
\begin{eqnarray}
\p_{k,DPHAV}(\langle n \rangle)&=&\displaystyle \frac{A^{k}e^{-A}}{n!}\sum_{h=0}^k
{k \choose h}
 \frac{\left(-1\right)^h }{2\pi}\left(\frac{B}{A}\right)^h \nonumber\\[1ex]
 &\times&{}_1F_2 \left[\left\{\hbox{$\frac12$}+\hbox{$\frac12$}h\right\},
 \left\{\hbox{$\frac12$}+\hbox{$\frac12$}h\right\},
 \hbox{$\frac12$}+\hbox{$\frac14$}B^2 \right]\nonumber\\[1ex]
 &\times&\displaystyle\frac{\Gamma\left(\hbox{$\frac12$}+ \hbox{$\frac12$}h\right)
 \Gamma\left(\hbox{$\frac12$}\right)}{\Gamma\left(1+ \hbox{$\frac12$}h\right)}\ , \label{eq:phaseaver}
\end{eqnarray}
where $A = \alpha^2+|\beta|^2$, $B=2 \alpha |\beta|$ and $_1F_2(a,b,z)$ is the generalized hypergeometric function.
The distribution in Eq.~(\ref{eq:phaseaver}) has mean $\langle n \rangle=\alpha+|\beta|^2$ and variance $\sigma^{(2)}_{n}= \langle n \rangle\left({\cal K}\langle n \rangle+1 \right)$, with
${\cal K}\equiv 2\alpha^2|\beta|^2/(\alpha^2+|\beta|^2)^2$.

\section{Conditioning by PNR detectors}\label{s:cond:PNR}
The bipartite state $\varrho^{\rm (out)}(\alpha,\beta)$ we obtained by mixing the $\varrho_{\DPHAV}(\alpha,\beta)$ state with the vacuum state $\varrho_0 = | 0 \rangle\langle 0 |$ at a 50:50 BS can be written as (without loss of generality we can take $\alpha \in {\mathbbm R}$, $\alpha\ge 0$)
\begin{align}
\varrho^{\rm (out)}(\alpha,\beta) &=
U_{\BS} \varrho_{\DPHAV}(\alpha,\beta)\otimes \varrho_{0}U^{\dag}_{\BS}  \nonumber\\[1ex]
&= \int_{\mathbbm C} \d^2 z\, P(z - \alpha; \beta) \, |z/\sqrt{2} \rangle \langle z/\sqrt{2} |
\nonumber \\
& \hspace{1cm}\otimes |- z/\sqrt{2} \rangle \langle-z/\sqrt{2} | , \nonumber\\[1ex]
&= \int_{-\pi}^{\pi} \frac{d\phi}{2\pi}\,
| \a+\b\,e^{i \phi}\rangle\langle \a + \b\,e^{i \phi} | \nonumber \nonumber\\
& \hspace{1cm}\otimes
| -(\a+\b\,e^{i \phi})\rangle\langle -(\a + \b\,e^{i \phi}) |,
\end{align}
where $U_{\BS}$ is the unitary operator describing the action of the BS, $\a = \alpha/\sqrt{2}$ and $\b = |\beta|/\sqrt{2}$.
\par
\begin{figure*}[htbp]
\includegraphics[width=0.45\columnwidth]{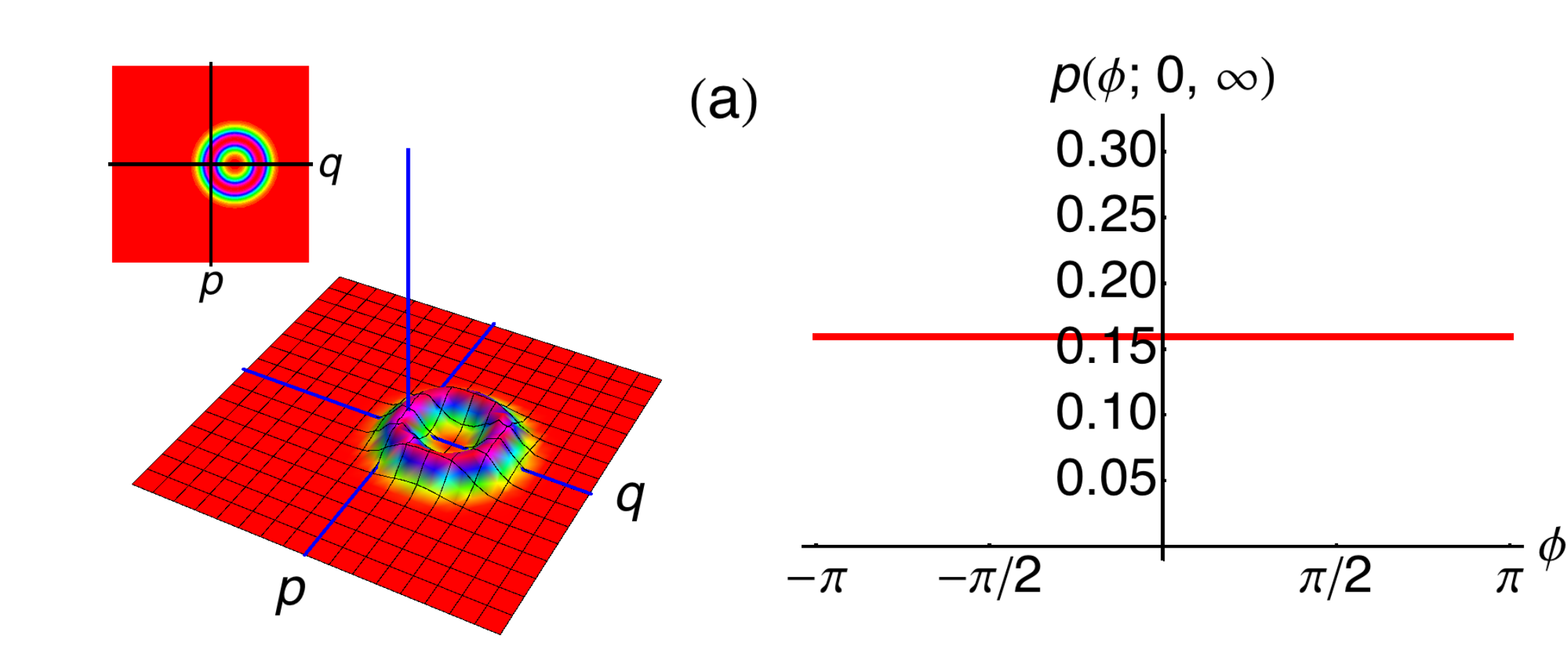}\hfill
\includegraphics[width=0.45\columnwidth]{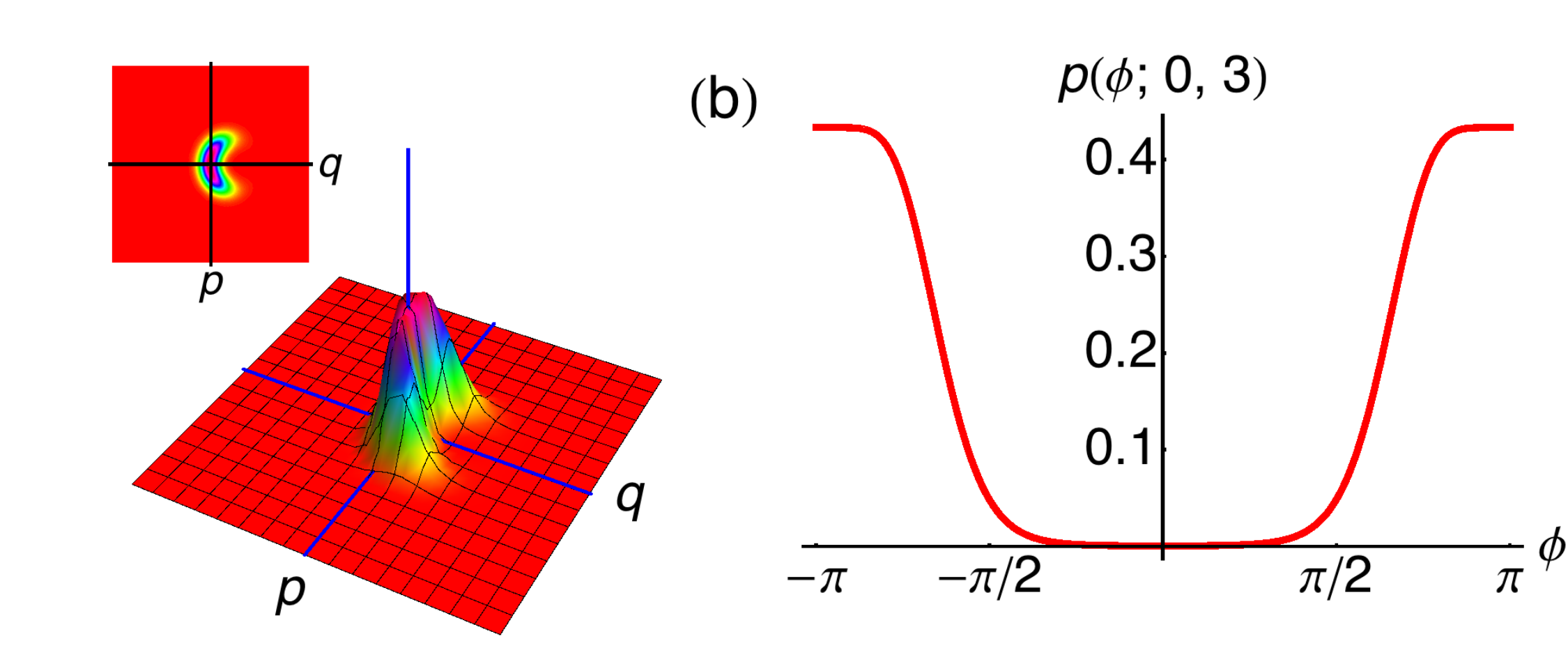}
\includegraphics[width=0.45\columnwidth]{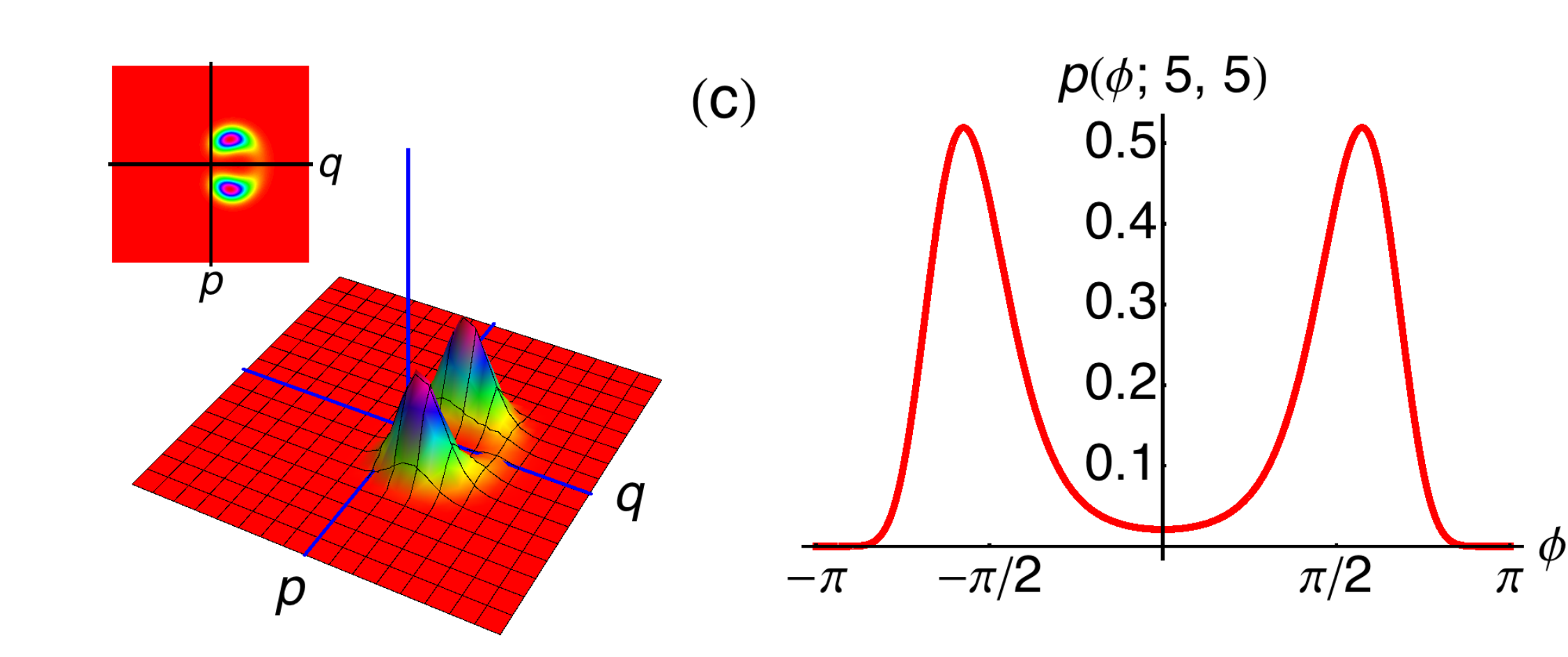}\hfill
\includegraphics[width=0.45\columnwidth]{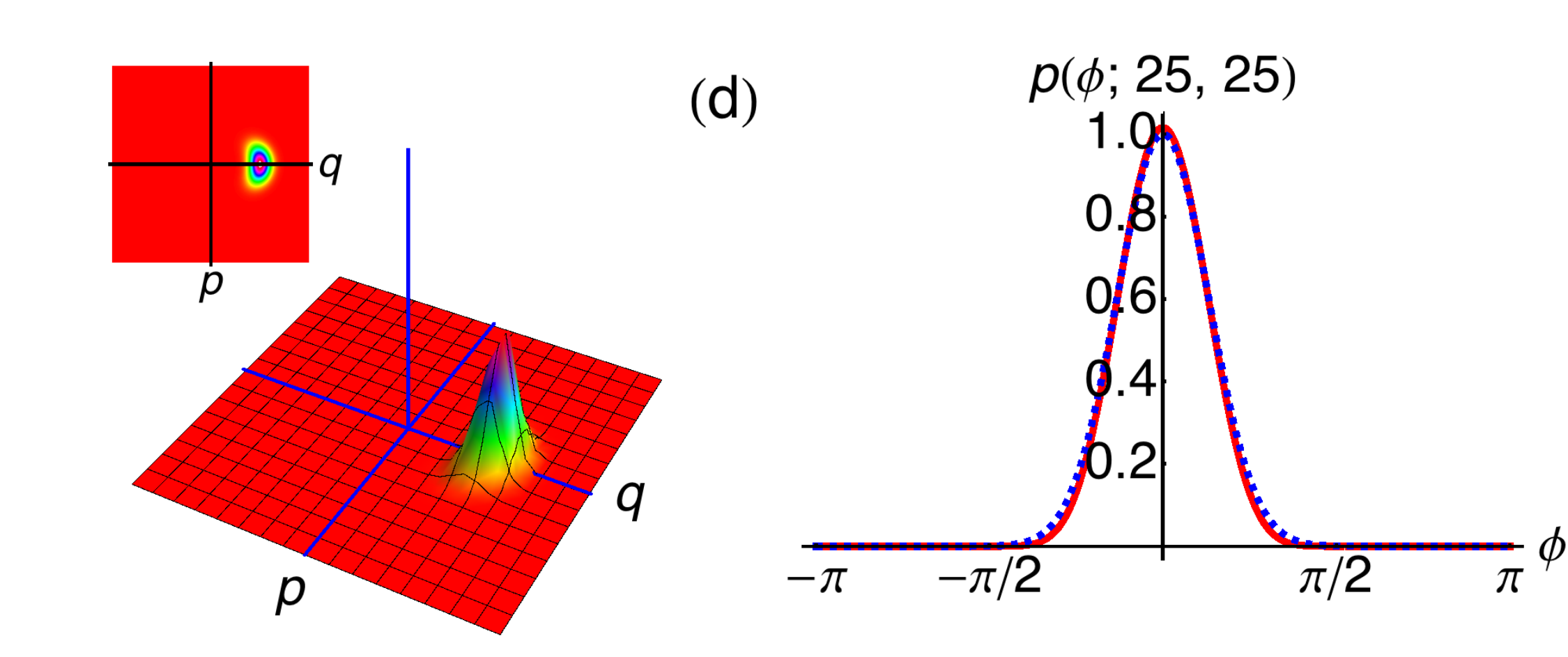}
\caption{(Color online) Plots of $p(\phi; k_1, k_2)$ for different values of $k_1$ and $k_2$ and of the corresponding Wigner function of the conditional state obtained starting from a DPHAV state. We set $\alpha = \sqrt{7}$ and $\beta=\sqrt{6}$. The dashed line in the plot (d) is the Gaussian approximation in Eq.~(\ref{gauss:app}) of the probability distribution $p(\phi; k_1, k_2)$ for $k_1=k_2 \gg |\alpha + \beta|^2/2 \approx 12.98$, which can be seen as a phase-diffusion process. \label{f:stat:W3D}}
\end{figure*}
As we perform PNR measurement on the reflected beam, which has a natural expansion in the photon-number basis, from now on we focus on the photon-number expansion of the states, whereas the $P$-representation can be directly obtained from our results.
In particular, it is worth noting that although the $P$-representation of the conditional states may be non-trivial, it is always positive, underlining the classical nature of the states \cite{glauber63}.
The expansion in the photon-number basis of the PNR measurement we are considering here is
\begin{equation}\label{PS:cond}
\Pi(k_1,k_2) = \sum_{h=k_1}^{k_2} | h \rangle\langle h |,
\end{equation}
with $0 \le k_1\le k_2$. If $k_1 = k_2 = k$, we have $\Pi(k,k) = | k \rangle\langle k |$ and the measurement subtracts $k$ photons from the input state. Indeed, if $k_1 \ne k_2$ we can generate a large family of conditional states. Therefore, the single-mode conditional state writes (for the sake of simplicity in the following we drop the explicit dependence on the amplitude $\a$ and $\b$)
\begin{align}\label{cond:state}
\varrho(k_1,k_2) &= \frac{1}{{\cal N}}
\int_{-\pi}^{\pi} \frac{d\phi}{2\pi}\,
| \a+\b\,e^{i \phi}\rangle\langle \a + \b\,e^{i \phi} | \nonumber\\
& \hspace{0.5cm} \times
\langle -(\a+\b\,e^{i \phi})| \Pi(k_1,k_2)| -(\a + \b\,e^{i \phi}) \rangle,
\end{align}
where we introduced the normalization factor
\begin{align}
{\cal N} &= {\cal N}(k_1, k_2)\nonumber\\[1ex]
&= \int_{-\pi}^{\pi} \frac{d\phi}{2\pi}\,
\langle -(\a+\b\,e^{i \phi})| \Pi(k_1,k_2)| -(\a + \b\,e^{i \phi}) \rangle
\nonumber  \\[1ex]
&= \sum_{h=k_1}^{k_2}\int_{-\pi}^{\pi} \frac{d\phi}{2\pi}\,
\p_h\left( \a^2+\b^2+2\a \b\cos\phi \right),
\label{cond:norm}
\end{align}
in which $\p_h$ is the Poissonian distribution defined in Eq.~(\ref{poisson}).
Equation~(\ref{cond:state}) can be also written as
\begin{align}
\varrho(k_1,k_2) =
\int_{-\pi}^{\pi} d\phi\,
p(\phi; k_1,k_2)\,
| \a+\b\,e^{i \phi}\rangle\langle \a + \b\,e^{i \phi} |
\label{cond:state:p}
\end{align}
where $p(\phi; k_1,k_2)$ is the probability distribution of the variable
$\phi$ given $k_1$ and $k_2$ (and, of course, $\alpha$ and $\beta$)
\begin{align}
p(\phi; k_1 , k_2) &= p(\phi; \a,\b; k_1 , k_2) \nonumber\\
&= \frac{\langle -(\a+\b\,e^{i \phi})| \Pi(k_1,k_2)| -(\a + \b\,e^{i \phi}) \rangle}
{2\pi\, {\cal N}}, \nonumber  \\
&=\frac{\sum_{h=k_1}^{k_2} \p_h\left( \a^2+\b^2+2\a \b\cos\phi \right)}{2\pi {\cal N}}.
\end{align}
From Eq.~(\ref{cond:state:p}) it follows that we can engineer different kinds of statistical mixtures of coherent states by suitably selecting the values of $k_1$ and $k_2$. In Fig.~\ref{f:stat:W3D} we plot the probability distribution $p(\phi;  k_1 , k_2)$ and the corresponding Wigner function of the conditional state for different choices of $k_1$ and $k_2$ in the case of a DPHAV state with $\alpha = \sqrt{7}$ and $\beta=\sqrt{6}$.
\par
In particular, we can identify two relevant cases:
\begin{itemize}

\item $k_1=0$ and $k_2 \to \infty$: we have the identity operator $\Pi(0,\infty) = \Id$ and $p(\phi; 0, \infty) = (2\pi)^{-1}$, and we obtain a DPHAV state with half the energy with respect to the input one due to the presence of the 50:50 BS [see Fig.~\ref{f:stat:W3D} (a)].

\item $k_1 = k_2 = k$: now we obtain $\Pi(k,k) = | k \rangle \langle k |$, $i.e.$, the projector onto the photon-number state $| k \rangle$, and we find
\begin{align}
p(\phi; k,k) &= \frac{\exp[-(\a^2+\b^2+2\a \b \cos\phi)]}{2\pi {\cal N}}\nonumber\\
&\hspace{0.5cm} \times\frac{(\a^2+\b^2+2\a \b \cos\phi)^k}{k!}.
\end{align}
In general for $k < (\a+\b)^2 $ the probability $p(\phi; k,k)$, which is indeed a non-Gaussian distribution, is double peaked in the interval  $[-\pi,\pi]$ [see Fig.~\ref{f:stat:W3D} (c)] and the two maxima occur at the phase values
\begin{equation}
\phi^{\rm (max)}_{\pm} = \pm \arccos \left[
1-\frac{(\a + \b)^2-k}{2\a\b}
\right].
\end{equation}
For $k \ge (\a+\b)^2$ we have only one peak at $\phi^{\rm (max)}=0$ and, in particular, for $k \gg (\a+\b)^2$, the probability distribution reduces to the normal distribution
\begin{equation}\label{gauss:app}
p(\phi; k,k) \approx \frac{1}{\sqrt{2\pi \Delta^2}}
\exp{\left(-\frac{\phi^2}{2\Delta^2}\right)},
\end{equation}
with (remember that we have $\a, \b \ge 0$)
\begin{equation}
\Delta^2 = \frac{(\a+\b)^2}{2\a \b \left[k - (\a+\b)^2\right]},
\end{equation}
as shown in Fig.~\ref{f:stat:W3D} (d). This last case is formally analogous to that of a coherent state undergoing a phase-diffusion process \cite{PDiff:geno}.

\end{itemize}

One of the properties of the states $\varrho(k_1,k_2)$ written in Eq.~(\ref{cond:state:p}), for fixed displacement amplitude $\a$ and PHAV state amplitude $\b$, is the symmetry of their photon distributions with respect to the exchange $\a \leftrightarrow \b$, namely
\begin{eqnarray} \label{p:m}
p_n(\a,\b;k_1,k_2)  &=& p_n(\b,\a;k_1,k_2)\nonumber\\
&=& \langle n | \varrho(k_1,k_2) | n \rangle \equiv p_n(k_1,k_2)
\end{eqnarray}
This feature makes it impossible to distinguish the displacement amplitude from the PHAV state one by means of a direct detection scheme, that is a scheme involving only PNR detectors.
\par
On the contrary, the non-Gaussianity of $\varrho(k_1,k_2)$ strongly depends on the value of the PHAV state amplitude $\b$ and becomes different by exchanging $\a\leftrightarrow\b$.
There are different ways to assess the non-Gaussianity of a state $\varrho$. Here we consider the relative entropy of non-Gaussianity \cite{ng:geno:08}. Given a generic state $\varrho$, this quantity is defined as the difference between the von Neumann entropy $S(\sigma)= -\hbox{Tr}[\sigma\ln \sigma]$ of a reference Gaussian state $\sigma$ and that of the state $\varrho$ under investigation, namely
\begin{equation}\label{nonG:r:e}
\delta(\varrho) = S(\sigma) - S(\varrho).
\end{equation}
The reference state $\sigma$ is a Gaussian state chosen to have the same mean value and covariance matrix as the state $\varrho$ \cite{ng:geno:08}, namely
\begin{align}
\langle x_\theta \rangle_{\varrho} &= \langle x_\theta \rangle_{\sigma} \quad \forall\theta \\
\Delta_{\varrho}^2(x) = \Delta_{\sigma}^2(x) &,\quad
\Delta_{\varrho}^2(y) = \Delta_{\sigma}^2(y) \\
\langle [x,y]_{+}\rangle_{\varrho} - 2\langle x \rangle_{\varrho}\langle y \rangle_{\varrho} &=
\langle [x,y]_{+}\rangle_{\sigma} - 2\langle x \rangle_{\sigma}\langle y \rangle_{\sigma},
\end{align}
in which $\Delta_A^2 (X) = \langle (X - \langle X \rangle_{A})^2 \rangle_{A}$,
 $\langle \cdots \rangle_{A} = \mbox{Tr}[\cdots \,A]$, $[x,y]_{+} = xy+yx$ and
\begin{equation}
x_\theta = \frac{a^\dag\,e^{i\theta}+a\,e^{-i\theta}}{\sqrt{2}},
\end{equation}
is  the quadrature operator with $x\equiv x_0$ and $y \equiv x_{\pi/2}$. In the case of the states $\varrho=\varrho(k_1,k_2)$ we have
\begin{align}
\Delta_{\varrho}^2(x) = \frac12 &+ 2b^2
\int_{-\pi}^{\pi} d\phi\,p(\phi;k_1,k_2)\,\cos^2\phi \nonumber \\
&- 2b^2\left[\int_{-\pi}^{\pi} d\phi\,p(\phi;k_1,k_2)\,\cos\phi\right]^2,\\
\Delta_{\varrho}^2(y) = \frac12 &+ 2b^2
\int_{-\pi}^{\pi} d\phi\,p(\phi;k_1,k_2)\,\sin^2\phi,
\end{align}
and $ \langle [x,y]_{+}\rangle_{\varrho} - 2\langle x \rangle_{\varrho}\langle y \rangle_{\varrho} = 0$.
As one may expect from the classicality of the states and from the choice of the parameters, we have $\Delta_{\varrho}^2(y) \ge \Delta_{\varrho}^2(x) \ge 1/2$, which means that both the quadrature variances cannot be below the shot noise. Indeed, the behavior of $p(\phi; k_1,k_2)$ leads to statistical mixtures of coherent states with a non-Gaussianity that strongly depends on the particular choice of $k_1$ and $k_2$.

\section{Experimental setup}\label{s:exp}

\begin{figure}[htbp]
\centering\includegraphics[width=0.5\columnwidth]{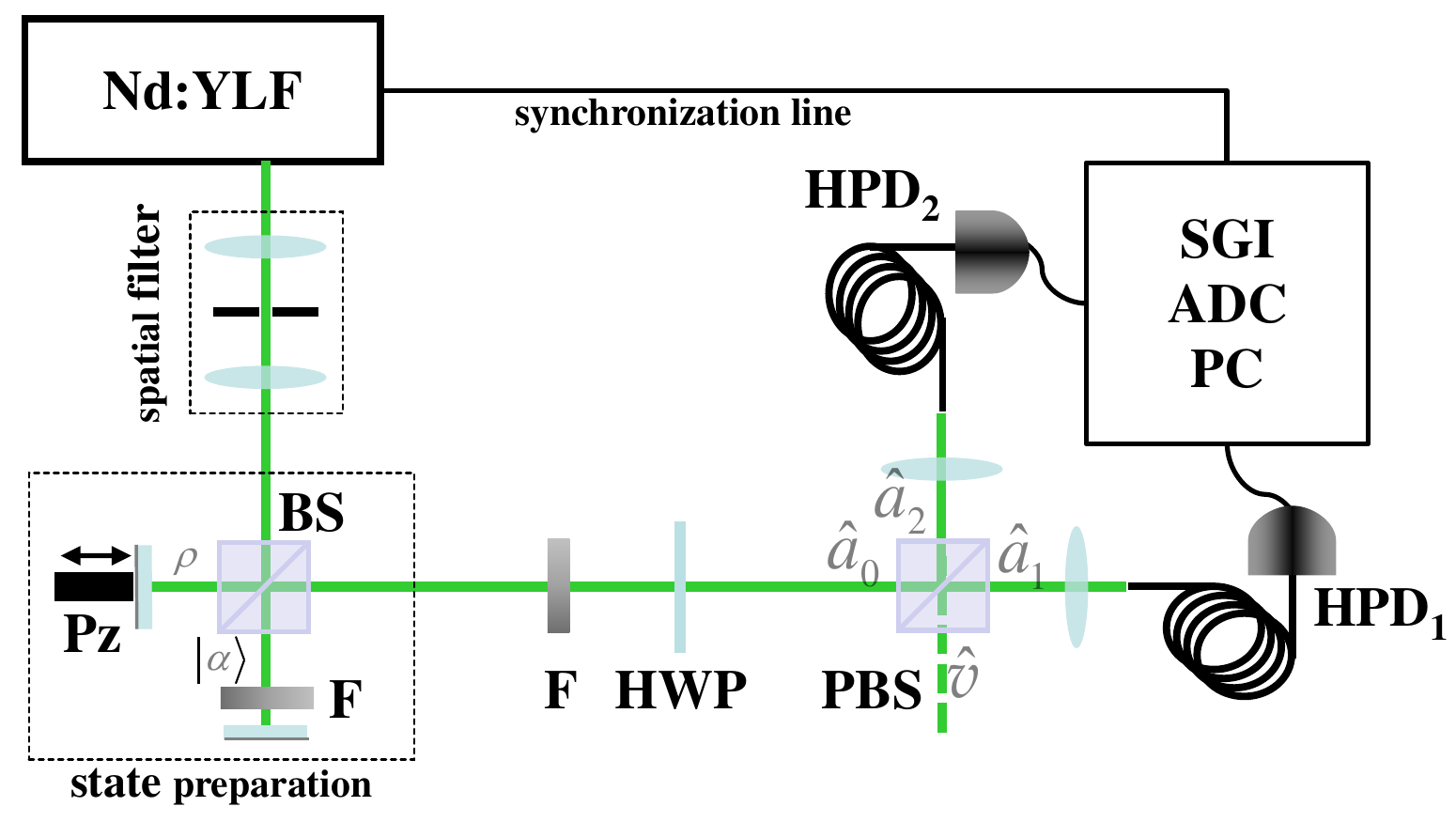}
\caption{(Color online) Experimental setup. F$_j$: variable neutral density filter; BS: 50:50 beam splitter; Pz: piezoelectric movement; MF: multimode fiber (600 $\mu$m core); HWP: half-wave plate; PBS: polarizing cube beam-splitter; HPD: hybrid photodetectors.} \label{f:setup}
\end{figure}

The experimental setup we used to produce DPHAV states is shown in Fig.~\ref{f:setup}.
The DPHAV state is obtained by sending the second-harmonic pulses ($\sim$5.4~ps, 523~nm) of a Nd:YLF mode-locked laser amplified at 500 Hz (High Q Laser Production) into a Michelson interferometer (see dotted box in Fig.~\ref{f:setup}). The mirror located in the reflected arm of the interferometer is mounted on a piezoelectric movement, whose displacement is operated at a frequency of 100 Hz and covers a 12 $\mu$m span in order to produce the PHAV state. The beam in the transmitted arm is kept coherent and reflected back to superimpose to the PHAV state: at the output of the beam splitter we have a DPHAV state. On both arms variable neutral-density filters are inserted to adjust the values of the PHAV state and of the displacement independently.
The DPHAV state is then sent to a second beam splitter whose outputs are collected by two multimode fibers and delivered to a pair of hybrid photodetectors (HPD, model R10467U-40,
maximum quantum efficiency $\sim0.5$ at 500~nm), which act as PNR detectors.
According to the strategy extensively described in Refs.~\cite{JMO09}, the experimental data, given in terms of output voltages, are processed in a self-consistent way, without any a-priori calibration of the detection chain and any background subtraction, and converted in numbers of detected photons. In this way we are able to reconstruct the statistics of detected photons and to calculate shot-by-shot detected-photon correlations.
\par
Due to the non unit quantum efficiency $\eta$ of the PNR detectors, there is a difference between the incident number of photons and the number of detected photons. In this last case, the projector on the photon-number basis, i.e., $|k\rangle \langle k|$ should be replaced as follows \cite{FOP:05}
\begin{equation}\label{proj:eta}
|k\rangle \langle k| \to \Theta_k (\eta) =
\sum_{s=k}^{\infty} {s \choose k}\,
\eta^{s}(1-\eta)^{s-k} \, |s \rangle \langle s|.
\end{equation}
We remark that all these results have been obtained in terms of photons, but actually they are also valid for detected photons because we are considering classical states, which are invariant under Bernoullian detection.
Thus the statistical properties do not change and the effects of the substitution in Eq.~(\ref{proj:eta}) are just a rescaling of the amplitudes, i.e., $|\alpha|^2 \to \eta |\alpha|^2$  and $|\beta|^2 \to \eta |\beta|^2$.
\par
From now on, we will refer to the detected number of photons $m$, if necessary with suitable subscripts.
\begin{figure}[htbp]
\centering\includegraphics[width=0.5\columnwidth]{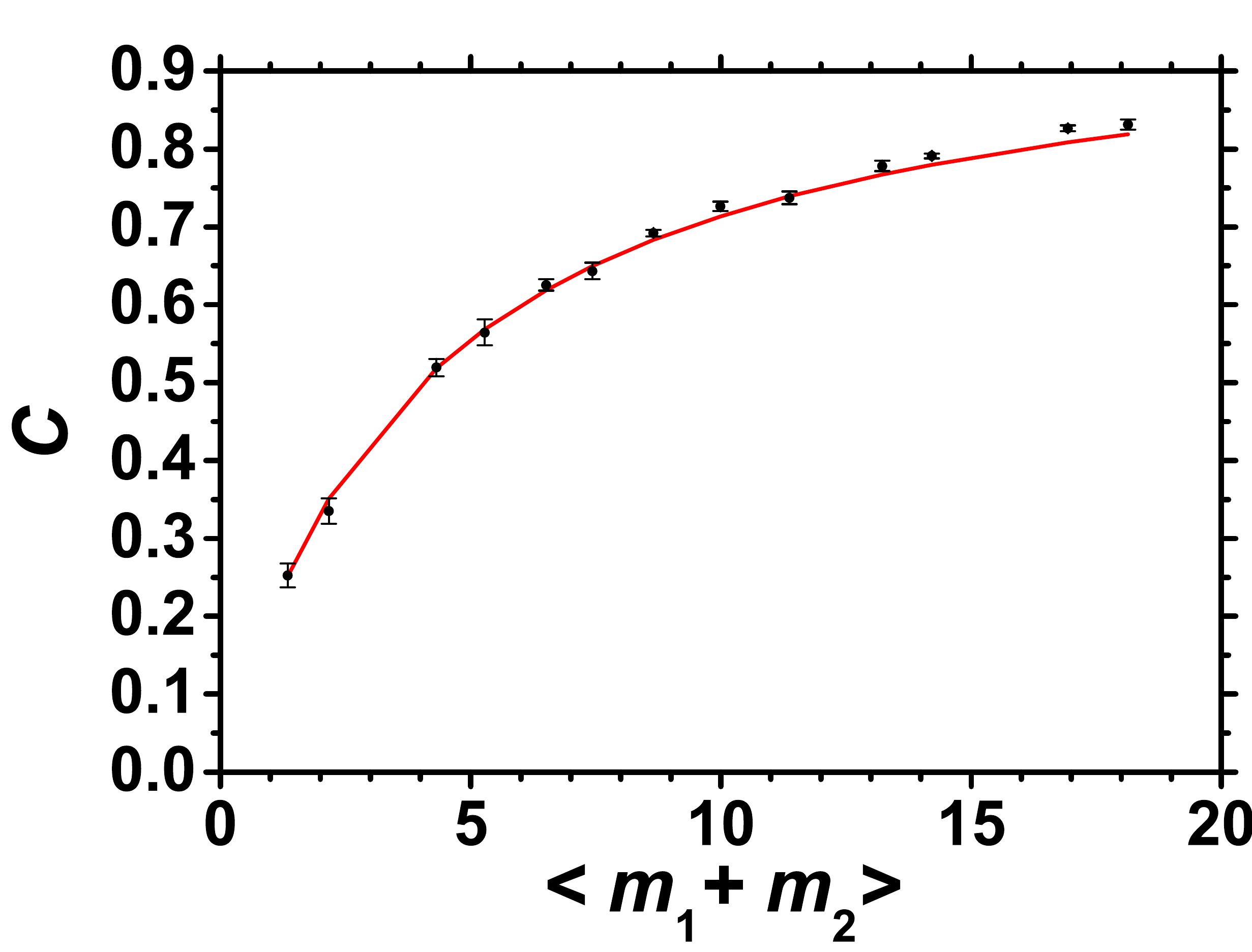}
\caption{(Color online) Second-order correlation coefficient as a function of the overall mean number of detected photons at the outputs of the beam splitter. Dots: experimental data; line: theoretical expectation evaluated in the experimental parameters.} \label{corrDPHAV}
\end{figure}
We start our analysis by investigating the detected-photon correlations between the two beams produced by splitting a DPHAV state at a 50:50 BS. The correlation coefficient between the outputs of the BS depends only on the input amplitudes $\alpha^2$ and $|\beta|^2$. According to Eq.~(\ref{gamma}) in which we insert mean values and variance of the PHAV state, we have
\begin{equation}\label{eq:correl}
\GGamma \equiv
\GGamma(\alpha^2,|\beta|^2) = \frac{\alpha^2\left|\beta\right|^2}{\alpha^2 + \left|\beta\right|^2
+\alpha^2\left|\beta\right|^2}.
\end{equation}
Note that $\GGamma(\alpha^2,|\beta|^2) = \GGamma(|\beta|^2,\alpha^2)$: as we mentioned in the previous Section, direct detection leads to quantities which are symmetric with respect to the PHAV state and displacement amplitudes. Figure~\ref{corrDPHAV} shows the experimental behavior of the correlation coefficient together with the theoretical expectation obtained by using the experimental parameters, determined in a self-consistent way as described in \cite{PRA12}, in Eq.~(\ref{eq:correl}). As anticipated in the Introduction, the existence of correlations between the two emerging beams makes the conditional PS process possible.

\section{Conditional states}\label{s:cond:s}

The conditional states $\varrho(k_1,k_2)$ are obtained by conditioning a DPHAV state $\varrho_{\DPHAV}(\alpha,\beta)$ divided at a 50:50 BS according to the projector defined in Eq.~(\ref{PS:cond}) and the substitution in Eq.~(\ref{proj:eta}). First of all, we measure the  photon-number statistics of the conditional states $p_{m}(k_1,k_2)$.
\begin{figure}[htbp]
\centering\includegraphics[width=0.5\columnwidth]{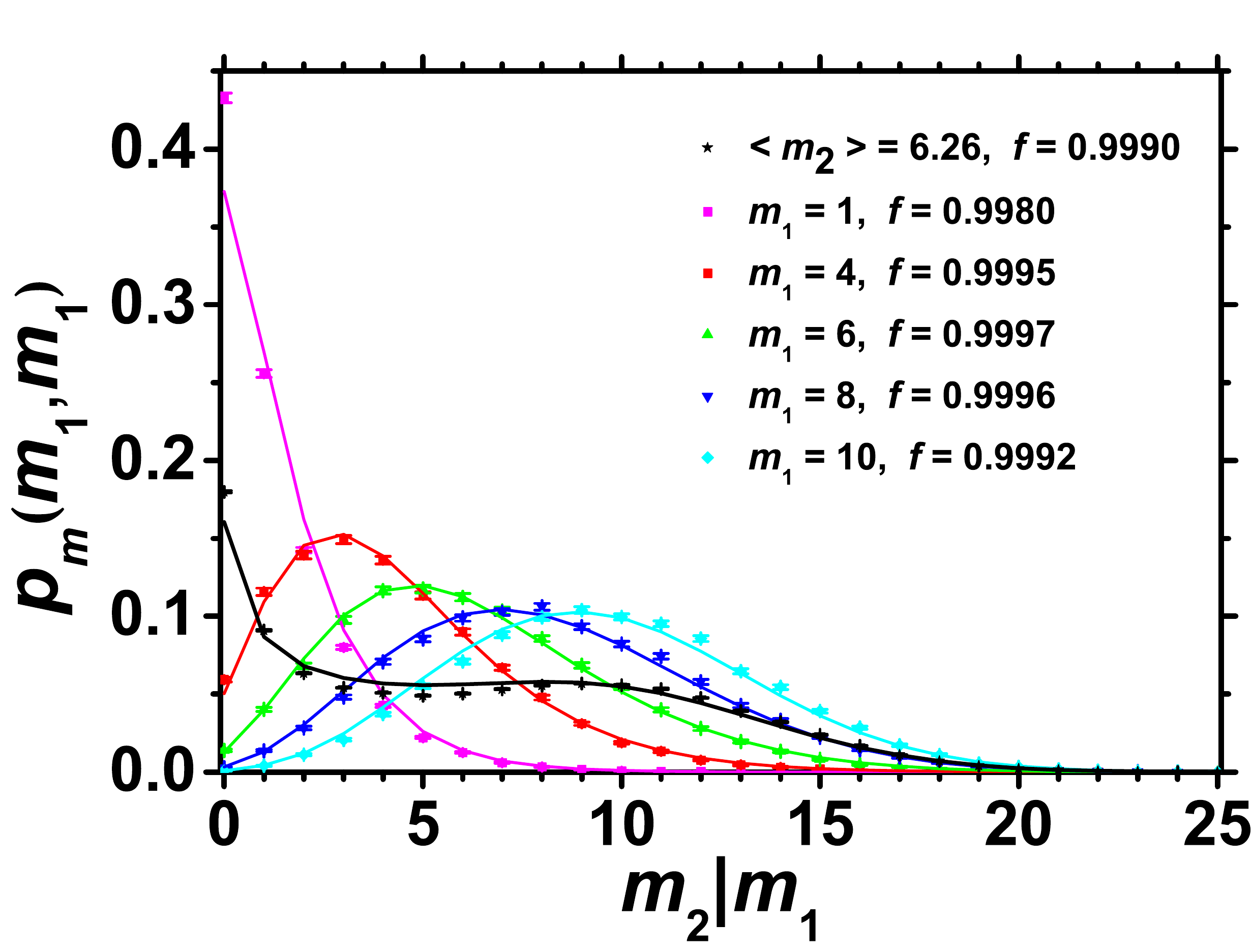}
\caption{(Color online) Detected-photon distributions of the conditional states obtained from a DPHAV state having $\alpha^2=6.17$ and $|\beta|^2=7.13$ for the condition ``$=m_1$''. Symbols: experimental data; lines: theoretical expectations. The unconditional state is also shown in black.} \label{istoCONDeq}
\end{figure}
In Fig.~\ref{istoCONDeq} we show the experimental photon-number distributions of the conditional states (dots) obtained by selecting a precise value of $k_1=k_2=m_1$ (condition ``$=m_1$'').  The theoretical expectations of Eq.~(\ref{p:m}) (written in terms of detected photons) are superimposed to the data. Similar results can be obtained for the other conditions. The good quality of our data is certified by the high values of the fidelity evaluated as $f=\sum_m\sqrt{p^{exp}_m p^{th}_m}$.
\par
In order to experimentally verify the symmetry exhibited by the photon-number distributions in Eq.~(\ref{p:m}), we consider two input DPHAV states with the displacement and PHAV state amplitudes exchanged, namely $\varrho_{\DPHAV}(\alpha,\beta)$ and $\varrho_{\DPHAV}(\beta,\alpha)$.
\begin{figure}[htbp]
\centering\includegraphics[width=0.5\columnwidth]{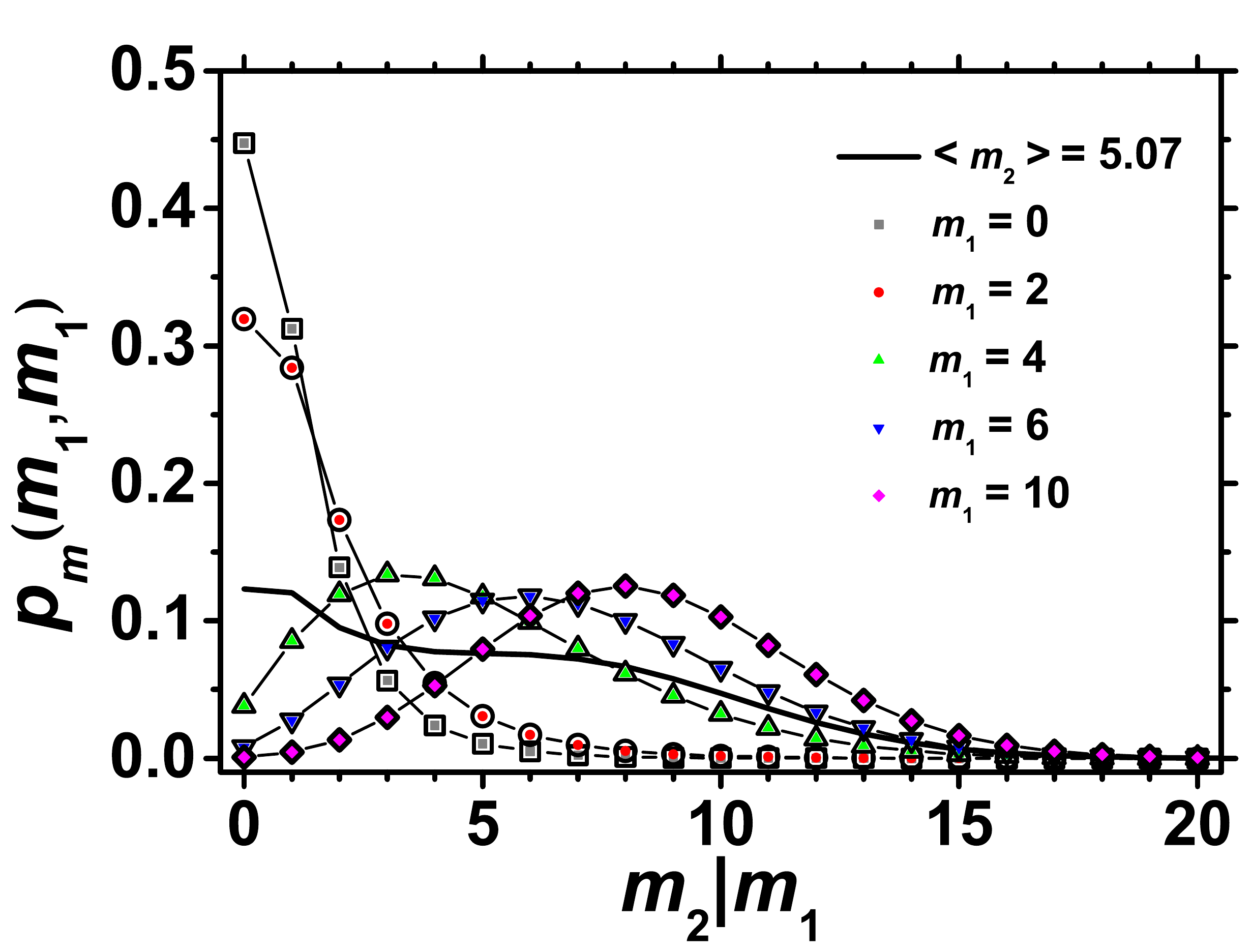}
\caption{(Color online) Detected-photon distributions of the conditional states obtained for the condition ``$=m_1$''. Full symbols: $\alpha^2=3$ and $|\beta|^2=7.13$; empty symbols: $\alpha^2=7.13$ and $|\beta|^2=3$. The two sets of histograms are indistinguishable.} \label{istoCONDsymm}
\end{figure}
The results are shown in Fig.~\ref{istoCONDsymm}, where we plot the experimental distributions obtained by imposing the condition $k_1=k_2=m_1$: as expected, the two situations are indistinguishable. Similar results are obtained also for other choices of $k_1$ and $k_2$, confirming our calculations.
\begin{figure}[htbp]
\centering\includegraphics[width=0.5\columnwidth]{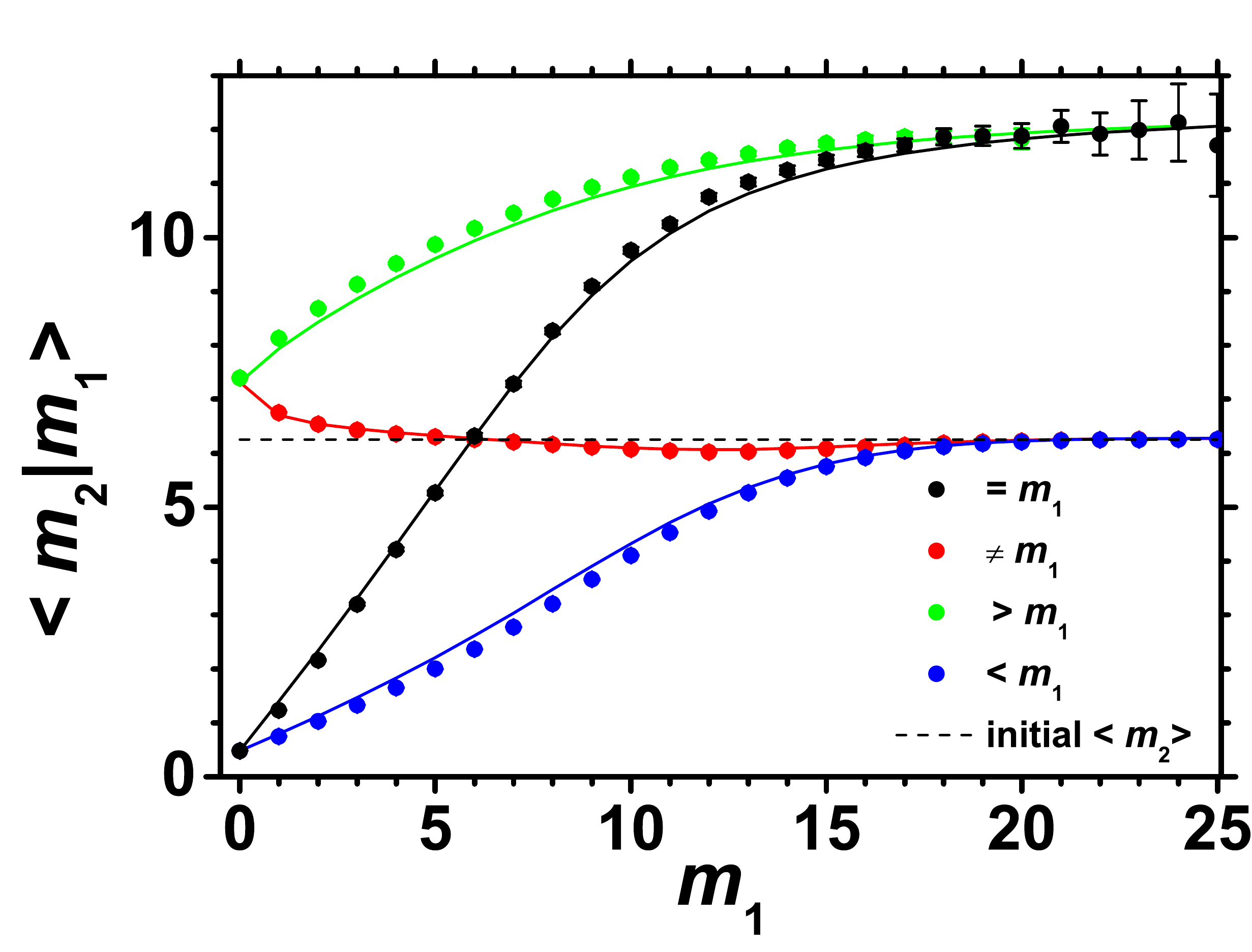}
\caption{(Color online) Mean values of conditional states as a function of the conditioning value obtained for the same parameters as in Fig.~\ref{istoCONDeq}. Symbols: experimental data; lines: theoretical expectations. The mean value of the initial state is displayed as dashed line. The error bars are smaller than the symbol size.} \label{meanCOND}
\end{figure}
Finally, in Fig.~\ref{meanCOND} we show the mean values of the conditional states as a function of the conditioning value.

\section{Non-Gaussianity}\label{s:nonG}

To quantify the resources of an optical state to be used in quantum information protocols, we can exploit the fact that in general a state is characterized by several parameters. The case of the DPHAV state is interesting because it can be described by its mean number of photons, the ratio between PHAV state and displacement in the original DPHAV state and the value of non-Gaussianity, which depends also on the conditioning value. We can operate on all these parameters independently: the amount of non-Gaussianity depends on PHAV state, while the overall mean value of the state is also influenced by the displacement.
\par
As we observed in Sec.~\ref{s:cond:PNR}, the non-Gaussianity of a DPHAV state and of its conditional counterparts strictly depends on the amplitude of the original PHAV state, that is the source of non-Gaussianity.
\par
The amount of non-Gaussianity of the DPHAV state in Eq.~(\ref{dm:D:phav}) is equal to that of the original PHAV state in Eq.~(\ref{dm}), being the displacement operation a Gaussian operation. As the PHAV state has a diagonal density matrix, the resulting expression of $\delta$ is particularly simple and only depends on its average number of photons $\langle n \rangle = |\beta|^2$ \cite{PHAV:rev}
\begin{align}\label{eq:nonGdiag}
\delta (\varrho_{\DPHAV})&= \delta (\varrho_{\PHAV})\nonumber\\
&=\sum_{k=0}^{\infty}\Bigg\{-\frac{\langle n \rangle^k}{(\langle n \rangle+1)^{k + 1}} \ln\left[\frac{\langle n \rangle^k}{(\langle n \rangle+1)^{k + 1}}\right]\nonumber\\
&\hspace{1.0cm}+ \p_k(\langle n \rangle) \ln\p_k(\langle n \rangle) \bigg\}.
\end{align}
It is worth pointing out that even in the case of diagonal states, we are not able to directly measure $\delta$, since Eq.~(\ref{eq:nonGdiag}) involves the distribution of the incident number of photons, whereas we have experimentally access only to detected photons. Nevertheless, we can define a lower bound $\varepsilon$ for the non-Gaussianity, $\varepsilon(\varrho)<\delta(\varrho)$, which is formally equal to Eq.~(\ref{eq:nonGdiag}) but is based on the detected-photons statistics \cite{IPS}.
\par
\begin{figure}[htbp]
\centering\includegraphics[width=0.5\columnwidth]{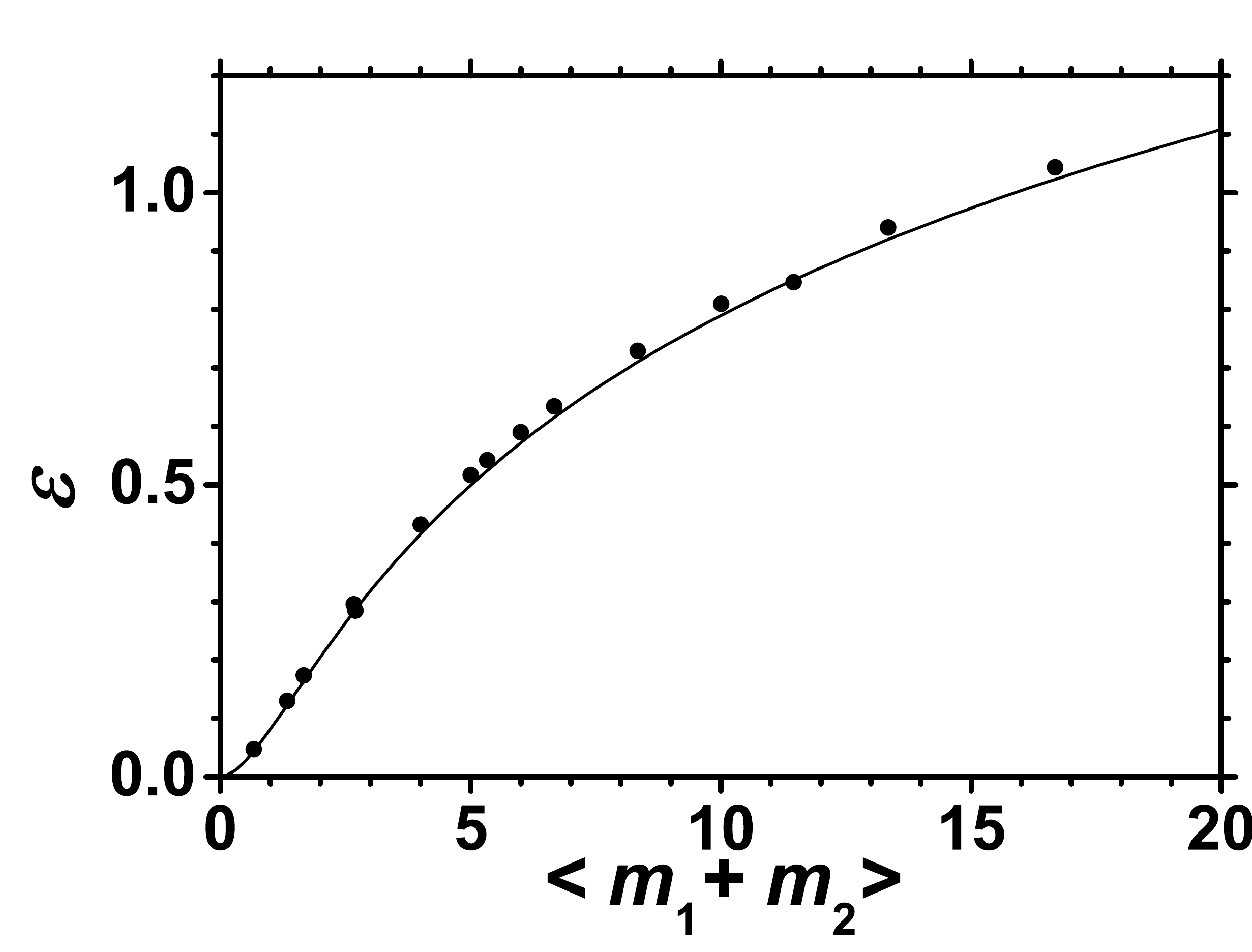}
\caption{(Color online) Non-Gaussianity measure for DPHAV states with $\alpha^2 = |\beta|^2$ as a function of their mean number of detected photons $<m>=\alpha^2+|\beta|^2$ in the state (symbols) along with theoretical values calculated according to Eq.~(\ref{eq:nonGdiag}) (line).} \label{nGphav}
\end{figure}

First of all we demonstrate that a Gaussian operation, that is the displacement, does not modify the non-Gaussianity of a PHAV state. Therefore, we calculate the density matrix of the DPHAV state from prime principles and compare the values of the calculated $\varepsilon$ with those obtained by measuring the PHAV state statistics.
In Fig.~\ref{nGphav} we plot the non-Gaussianity of a DPHAV state with $\alpha^2= |\beta|^2$ as calculated from the experimental statistics of detected photons for different values of the total energy (symbols). In the same figure we show the theoretical values of the PHAV state obtained from Eq.~(\ref{eq:nonGdiag}) (line).
\par
\begin{figure}[htbp]
\centering\includegraphics[width=0.5\columnwidth]{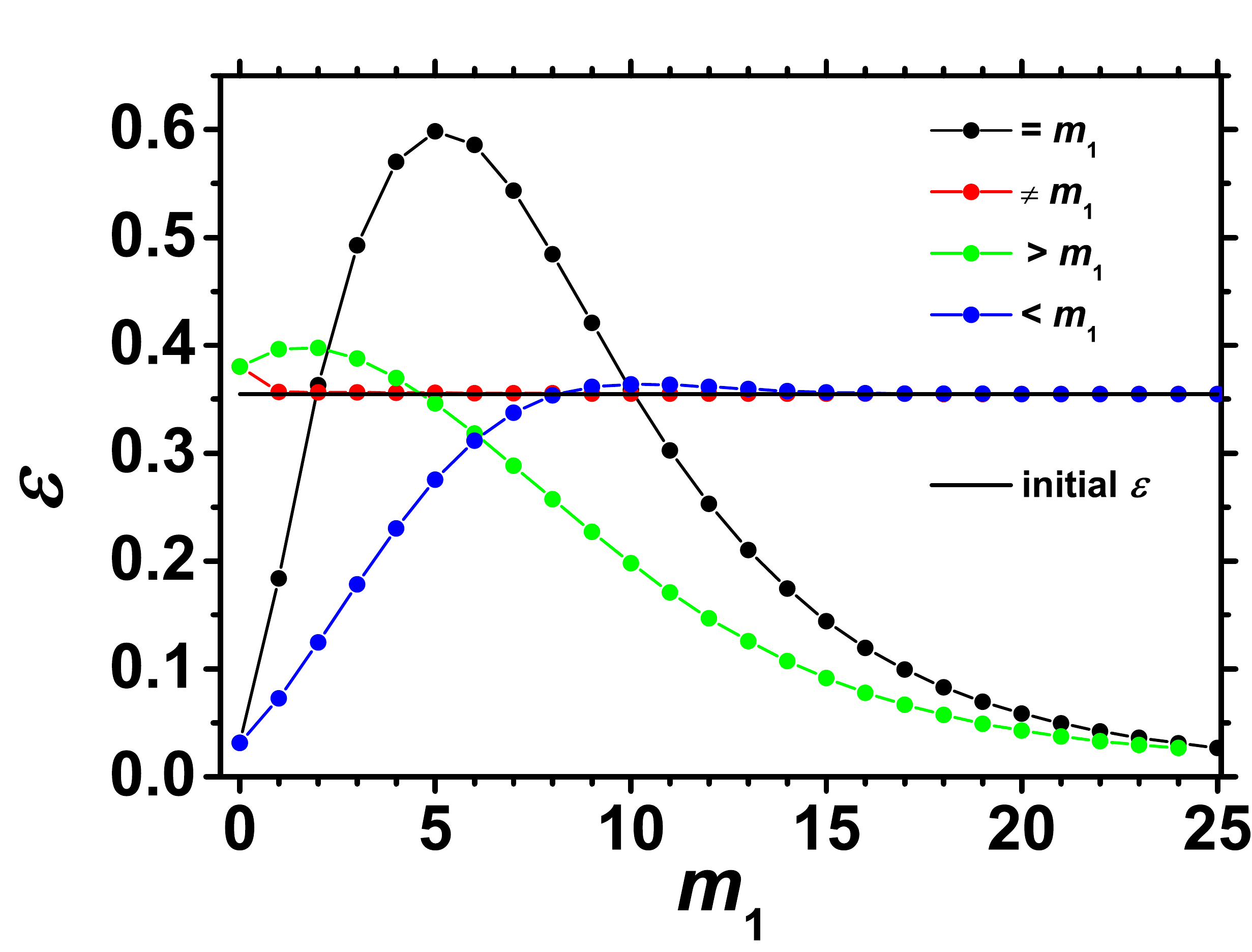}
\caption{(Color online) Values of the non-Gaussianity amount of the conditional states, as a function of the conditioning value, generated according to different rules: ``$=m_1$'' (black dots), ``$\neq m_1$'' (red dots), ``$> m_1$'' (green dots) and ``$\leq m_1$'' (blue dots). The data for the first conditions correspond to the histogram in Fig.~\ref{istoCONDeq}.} \label{ngDPHAV}
\end{figure}

As a further investigation of the contribution of the different experimental parameters to the amount of non-Gaussianity of the conditional states we study the lower bound $\varepsilon$ as a function of the conditioning value for a fixed choice of the mean number of detected photons in the initial DPHAV state. In Fig.~\ref{ngDPHAV} we plot $\varepsilon[\varrho(k_1,k_2)]$ as a function of the conditioning value $m_1$ for four different criteria (or rules) adopted to generate the conditional states: ``$=m_1$'', ``$\neq m_1$'', ``$>m_1$'' and ``$\leq m_1$''. The results show that the values of $\varepsilon$ depend on the conditioning operation and that, against intuition, the effect of non-Gaussian operations applied to a non-Gaussian state can determine either larger or smaller values of non-Gaussianity.
\par
First of all, we observe that the condition ``$=m_1$'' produces conditional states having the maximum variation of $\varepsilon$ with respect to the initial values of non-Gaussianity (see horizontal line in Fig.~\ref{ngDPHAV}). In particular, the operation corresponding to the selection rule ``$m_1= 0$'' produces a state with $\varepsilon\rightarrow 0$, that is a quasi-Gaussian state, even starting from a non-Gaussian initial PHAV state. This is due to the fact that selecting according to ``$m_1=0$'' is the only Gaussian operation among the conditional ones.
\par
The second evident feature in Fig.~\ref{ngDPHAV} is the maximum in $\varepsilon$ for the condition ``$=m_1$'' at a given value of $m_1$. Such behavior can be understood by considering the analytical expression of the conditional state given in Eq.~(\ref{cond:state:p}). In fact, if $m_1$ is less than the energy of the input state, $p(\phi; k_1,k_2)$ exhibits two peaks, thus containing a larger amount of non-Gaussianity with respect to that of the unconditioned PHAV state. As $m_1$ increases and becomes larger than the energy of the input state, the non-Gaussianity decreases and approaches zero: in this case $p(\phi; k_1,k_2)$ becomes the normal distribution. Nevertheless, it is worth noting that the conditional state obtained even for $m_1 \gg 1$ is still non-Gaussian. The experimental behavior of the non-Gaussianity presented in Fig.~\ref{ngDPHAV} for the other conditioning choices can be explained in similar ways.
\begin{figure}[htbp]
\centering\includegraphics[width=0.5\columnwidth]{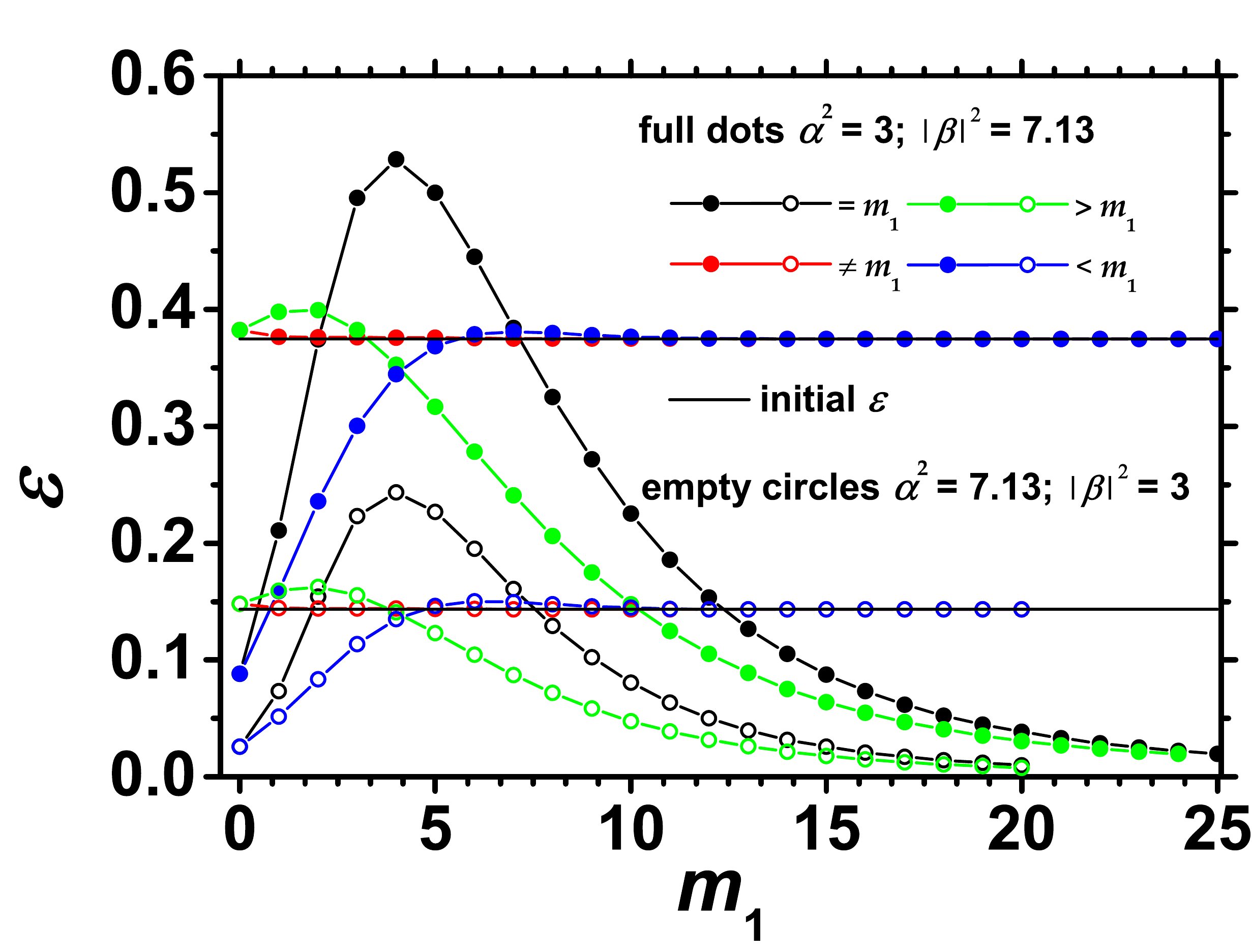}
\caption{(Color online) Values of the non-Gaussianity amount of the conditional states, as a function of the conditioning value, generated according to different rules: ``$=m_1$'' (black), ``$\neq m_1$'' (red), ``$> m_1$'' (green) and ``$\leq m_1$'' (blue). Full symbols: $\alpha^2=3$ and $|\beta|^2=7.13$; empty symbols: $\alpha^2=7.13$ and $|\beta|^2=3$.} \label{nGsymm}
\end{figure}
\par
In this context, it is also interesting to address the issue of the symmetry exhibited by the photon statistics $\p_{k,DPHAV}(\langle n \rangle)$ [see Eq.~(\ref{eq:phaseaver})] and by the photon-number distribution of the conditional states. In Fig.~\ref{nGsymm} we plot the values of the lower bound $\varepsilon$ for two symmetric situations $\alpha^2=3$ and $|\beta|^2=7.13$ and \emph{viceversa}. The behavior of the data is very similar for all the considered selection rules, but the absolute values are different. This confirms that the absolute value of the non-Gaussianity of the conditional states depends on the amount of displacement while the optimal choice of the conditioning value $m_1$ depends on the initial PHAV state.
\par
\begin{figure}[htbp]
\centering\includegraphics[width=0.5\columnwidth]{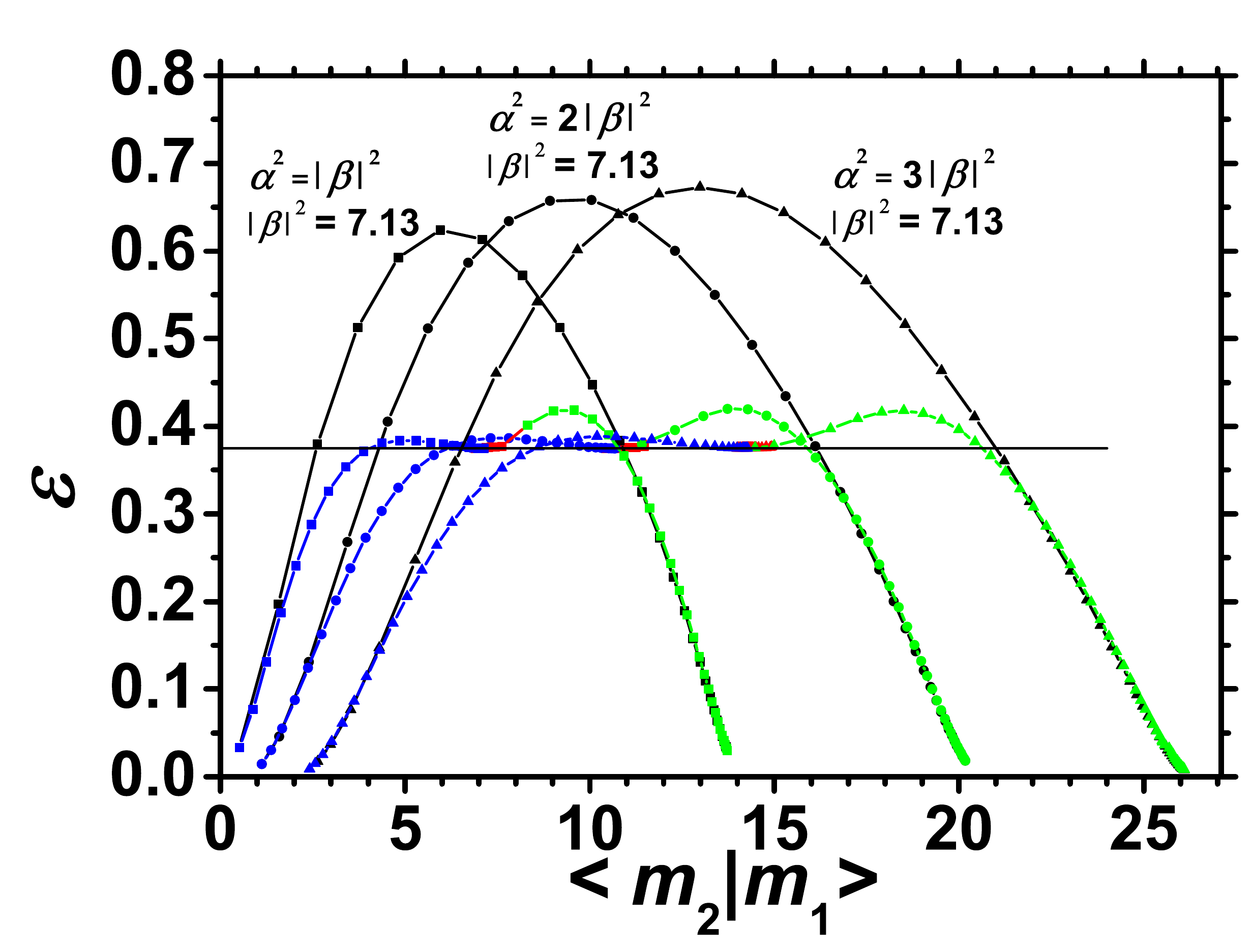}
\caption{(Color online) Values of the non-Gaussianity amount of the conditional states, as a function of their mean value, generated according to different rules (``$=m_1$'' (black), ``$\neq m_1$'' (red), ``$> m_1$'' (green) and ``$\leq m_1$'' (blue)) for different values of the displacement $\alpha^2=0.1,1,2,3 |\beta|^2$, at fixed $|\beta|^2=7.13$.} \label{ngALFA}
\end{figure}

The same conclusion can be reached from the insight of Fig.~\ref{ngALFA}, where we explore the effect on the amount of non-Gaussianity of changing the values of the displacement amplitude $\alpha$ by keeping the amplitude of PHAV state fixed. For the selection rule ``$=m_1$'', the maximum amount of non-Gaussianity is achieved for a mean value of the conditional state that depends on the overall energy in the original PHAV state. We note that, by virtue of Fig.~\ref{meanCOND}, the mean value of conditional state monotonically depends on the conditioning value $m_1$. We also note that the same values of non-Gaussianity can be obtained for different mean values. This suggests that we can independently tailor the value of non-Gaussianity and the mean value of the generated state by simply acting on the initial DPHAV state or by changing the conditioning operation and/or choosing a proper conditioning value.

\section{Discussion and Conclusions}\label{s:concl}

The results presented above show that the class of DPHAV states can be exploited to produce classical non-Gaussian states by performing multiple photon subtraction on DPHAV states divided at a beam splitter. The interesting parameters of the conditional states, that is their mean values and amount of non-Gaussianity, can be modified by choosing the initial mean value of the PHAV state and that of the displacement composing the DPHAV state and by properly selecting the conditioning value and the conditioning operation.
We demonstrated that all the properties that can be accounted by direct detection are invariant upon exchange of the role of displacement and phase-averaged component of the DPHAV state. Nevertheless the states are different because they are characterized by different amounts of non-Gaussianity, but this feature is somehow hidden in the internal structure of the state and cannot be revealed by direct detection measurements. To have access to the quantification of non-Gaussianity we thus need to perform phase-sensitive measurements able to reconstruct the Wigner function of the states or at least to recognize the coherent contribution in the state given by the displacement.

\section{Acknowledgments}
This work has been supported by MIUR (FIRB ``LiCHIS'' - RBFR10YQ3H). S.O. would like to acknowledge fruitful discussions with M.~G.~A.~Paris.

\end{document}